\documentclass[12pt]{svjour3}          
\smartqed  
\usepackage[margin=1in]{geometry}
\usepackage{setspace}
\usepackage{graphicx}
\usepackage{natbib}
\usepackage{xcolor}
\usepackage{amsmath}
\usepackage{mathrsfs}
\usepackage{amsfonts}
\usepackage{dsfont}
\usepackage{diagbox}
\usepackage{makecell}
\usepackage{color}
\usepackage{hyperref}
\usepackage{multirow}
\usepackage{mathtools}
\usepackage{comment}
\usepackage{mathtools}
\usepackage{booktabs}
\usepackage{float}

\usepackage{mathabx}
\usepackage{calligra}

\usepackage[perpage]{footmisc}

\DeclareMathOperator{\tf}{\mathfrak{t}}
\newcommand{\de}{\mathrm{d}}

\newcommand{\ee}{\mathds{E}}
\newcommand{\x}{{\mathbf x}}

\newcommand{\X}{\breve{X}}

\newcommand{\R}{\mathds R}

\newcommand{\LL}{\mathds{L}}
\newcommand{\rL}{r_{\LL}}
\newcommand{\1}{{\mathbf 1}}

\journalname{JABES}
\graphicspath{{./Figures/}}

\begin{document}

\title{
Rejoinder on `Marked spatial point processes: current state and extensions to point processes on linear networks' 
}
\author{
Matthias Eckardt$^{\text{1}}$ \and Mehdi Moradi$^{\text{2}}$ 
}

\authorrunning{
Eckardt and Moradi 
}
\institute{
Matthias Eckardt\\
\email{m.eckardt@hu-berlin.de}\\
ORCID: 0000-0002-4015-486X
\vspace{.1cm}\\
Mehdi Moradi (Corresponding author)\\
\email{mehdi.moradi@umu.se}\\
ORCID: 0000-0003-3905-4498
\\
\vspace{.1cm}
\at $^{\text{1}}$ Chair of Statistics, Humboldt-Universit{\"a}t zu Berlin, Berlin, Germany. \\  
\at
$^{\text{2}}$ Department of Mathematics and Mathematical Statistics, Ume\r{a} University, Ume\r{a}, Sweden
}

\date{Received: date / Accepted: date}

\maketitle

\begin{abstract}
We are grateful to all discussants for their invaluable comments, suggestions, questions, and contributions to our article. We have attentively reviewed all discussions with keen interest. In this rejoinder, our objective is to address and engage with all points raised by the discussants in a comprehensive and considerate manner. Consistently, we identify the discussants, in alphabetical order, as follows: CJK for Cronie, Jansson, and Konstantinou, DS for Stoyan, GP for Grabarnik and Pommerening, MRS for Myllymäki, Rajala, and Särkkä, and MCvL for van Lieshout throughout this rejoinder.

\end{abstract}

\section{Introduction}

In our article, we considered marked spatial point processes on two frequently examined state spaces, namely, $\R^2$ and linear networks; the latter increasingly got attention during the past two decades \citep{moradi2018spatial, BADDELEY2021100435}. Concerning marks, we focused on situations where the marks are qualitative, quantitative, or non-scalar, which we referred to as \textit{discrete and integer-valued marks}, \textit{real-valued}, and \textit{object-valued} marks. For practical applications, examples of object-valued quantities include the demographic evolution for different provinces of a country as in \citet[Figure 1]{Ghorbani2020} or the total crown defoliation per year for a set of trees as in \citet[Figure 4]{Eckardt2023MultiFunctionMarks}. In the province data example, the potential object of interest might be determining the heterogeneity/correlation of the curves across space, specifically whether nearby provinces exhibit greater similarity in their demographic evolution. For the tree example, a potential research question could involve determining the spatial correlation or the variation of total crown defoliation across a forest.      
Leaving point process models aside, given the constraints of space, we explicitly focused on mark summary characteristics, which play undeniable roles in marked point process analysis, for investigating and describing the (space-dependent) distributional properties of marks. Note that the empirical behavior of mark summary characteristics can be compared to an expected one under a null model, e.g. independent marking.

For marked spatial point processes in $\R^2$, we presented cross/dot-type summary characteristics, mark-weighted summary characteristics, various mark correlation functions, and frequency domain approaches. In the case of marked spatial point processes on linear networks, we proposed novel higher-order cross/dot-type summary characteristics, mark-weighted summary characteristics, and various mark correlation functions. The discussants have greatly enriched our work by bringing clarity to it and highlighting crucial points that have not been covered in our article, given the constraints of space. More specifically, CJK discussed mark spaces, product density functions, marked intensity reweighted stationarity, and general mark-weighted summary characteristics. DS paid the most attention to our article's applications and statistical interpretations. GP discussed cases where points are restricted to sub-spaces beyond the case of linear networks and pointed to practical applications as well as the consideration of compound mark functions instead of several single marks attached to a point. MRS thoroughly discussed mark point process models where the marks are either qualitative or quantitative. Finally, MCvL discussed cross/dot-type summary characteristics for multivariate random measures, asked about the limitations of mark correlation functions to stationary cases and IRMPS models for marked point processes on linear networks, and discussed non-parametric estimation of cross/dot-type higher-order summary characteristics and intensity functions when state space is $\R^2$. In this rejoinder, we provide further clarification on the points raised by the discussants.

\section{State spaces, stationarity, and restriction of the space}
Let $X=\{(x_i,m_i)\}_{i=1}^N,\ N < \infty, $ be a marked spatial point process with a spatial domain $S$ and a mark space $\mathcal{M}$, where, in our article, $S$ is either $\R^2$ or a linear network; both spatial and mark spaces are complete separable metric (csm) spaces. 
CJK discusses representative instances of the mark space $\mathcal{M}$, aligning closely with the content we presented in our article. We emphasize that, in all our considered cases, the mark space $\mathcal{M}$ is csm, and object-valued marks refer to non-scalar marks such as functions. 

Turning to stationarity-related concerns for marked point processes, GP adds that, based on their experience, inhomogeneity is not that common in nature, and incorporating it in the undertaken analysis may mask important interactions between events in question. In this regard, we agree with GP, especially in applications of forestry with small-sized observed windows. However, assessing this in broader contexts and other applications may pose challenges, particularly when dealing with a linear network as the state space; linear networks often inherently exhibit non-uniform structures, imposing additional complexity to the evaluation process. Some of the cross/dot-type inhomogeneous summary characteristics, such as $K$- and $J$-functions, regardless of state space, rely on the assumption that the underlying marked spatial point process $X$ is second-order intensity-reweighted stationary or intensity-reweighted moment stationary processes. For ease of notation and especially as, in our article, we did not discuss estimators of mark summary characteristics, we presented such properties for the unmarked version of $X$, denoted as $\X$. Nevertheless, CJK adequately presented the appropriate notation for such classes of spatial point processes in $\R^2$ and on linear networks in the framework of the marked spatial point process $X$, which shed more light on these concepts for interested readers.

MCvL asked if mark correlation functions such as those presented in Tables 1 and 3 of our article, which in practice are applicable in the case of real-valued marks, are inherently tied to stationarity. These summary characteristics, such as Stoyan's mark correlation function $\kappa_{mm}$, are, given a fixed interpoint distance $r$, simply functions of the marks and do not account for the spatial distribution of points.  In an inhomogeneous situation, for different interpoint distances $r$, the proportions of points satisfying $d(x,y)=r$ are influenced by the spatial distribution of points. Thus, one might need to take into account not only the mark distribution of nearby points but also the spatial distribution of the points in question. In an effort to do so, mark-weighted $K$-functions are proposed by \cite{pettinen1992forest} and \cite{Ghorbani2020} as mentioned by CJK in their discussion. In our article, we have also proposed mark-weighted $K$-functions for point processes on linear networks. However, the interpretation of mark-weighted summary characteristics might not be easy as different sources of variation exist, including pairwise interactions between points and between marks. 

Concerning IRMPS models for point processes on linear networks, \cite{cronie2020inhomogeneous} showed that Poisson processes are one example of such models. Additionally, they identified the essential conditions under which log-Gaussian Cox point processes on linear networks can be classified as IRMPS. In fact, they showed that for the log-Gaussian Cox point process $X$ on linear network $\LL$ with a random intensity measure $\Lambda(A)=\int_A \lambda(u) \de_1 u = \int_A \exp\{ Z(u)\} \de_1 u,\ A \subseteq \LL,$ where $Z$ is a Gaussian random field on $\LL$ with a mean function $\mu(u)$ and a covariance function $C(u,v),\ u,v \in \LL$, $X$ becomes IRMPS if 
\begin{align}\label{eq:cox}
    C(u_1,u_2)
    =
    \mathcal{C}(d_{\LL}(u_1,u),d_{\LL}(u_2,u)) \in \R, \quad u_1,u_2 \in \LL,
\end{align}
for any $u \in \LL$ and some function $\mathcal{C}$. In other words, the covariance function $C(u_1,u_2)$ should only depend on the distances between $u_1,u_2$ and an arbitrary point $u \in \LL$; these are recently extended to spatio-temporal settings by \cite{MoradiSharifi}. The development of IRMPS models is still in its early stages. Besides the information provided on log-Gaussian Cox point processes, to the best of our knowledge, there have been no additional findings in the literature regarding models that meet the IRMPS criteria or the fundamental conditions required for certain models to qualify as IRMPS. Thus, at present and in theoretical forms, our proposed higher-order summary characteristics can be calculated for Poisson processes and log-Gaussian Cox point processes under condition \eqref{eq:cox}. However, \cite{cronie2020inhomogeneous} and \cite{MoradiSharifi} showed that, in practice, their proposed $J$-functions are generally able to detect clustering/regularity/randomness in practice.

Lastly, GP addressed the spatial limitation beyond linear network settings by emphasizing practical scenarios where a collection of (disjoint) sub-spaces within a larger observation window is considered the state space. Practical examples they provided included nesting sites of birds on trees or rocks within a larger landscape, as well as the utilization of geostatistical covariates to impose spatial restrictions. These examples seem to us quite interesting; however, to the best of our knowledge, they have not received much attention within the literature, and 
determining how to proceed in such scenarios may not be immediately clear. As GP also noted, in these cases, interpreting the results may pose challenges, and in our opinion, additional insights from practical experts might be necessary.

\section{Mark filtering and non-parametric estimators}

Throughout our article, we did not discuss estimators for the summary characteristics we presented despite having estimated them in our applications. This was highlighted by CJK, GP, and MCvL. In particular, MCvL, in her discussion, presented non-parametric estimators for cross-type inhomogeneous nearest-neighbor distance distribution function $H_{ij}^{\mathrm{inhom}}(r)$ and intensity function $\lambda_i(x),\ x \in X_i$. Below, for marked point processes in $\R^2$, we briefly go through non-parametric estimators for some of the summary characteristics presented in our article. We also comment on estimators for the presented mark summary characteristics in settings where linear networks are the state space. 

The cross-type $K$-function, presented in equation (4) of our article, is usually estimated via
\begin{align}\label{eq:Kest}
    \widehat{K}_{ij}^{\mathrm{inhom}}(r)
    =
    \frac{1}{|W|}
    \sum_{x \in X_i}
    \sum_{y \in X_j}
    \frac{
    e(x,y) \1 \{ d(x,y) \leq r \}
    }{
    \lambda_i(x) \lambda_j(y)
    },
\end{align}
where $e(x,y)$ is en edge-correction and $|W|$ is the size of the window where the marked point process $X$ is observed \citep{Moller2004}. An estimator for the dot-type $K$-function can similarly be achieved. We add that these estimators can, in practice, be evaluated via functions \textsf{Kcross.inhom} and \textsf{Kdot.inhom} in the \textsf{R} package \textsf{spatstat} \citep{Baddeley2015}; few edge corrections are accessible \citep[Chapter 7]{Baddeley2015}. For cross-type inhomogeneous nearest-neighbor distance distribution function $H_{ij}^{\mathrm{inhom}}(r)$, given in equation (6) in our article, MCvL presented an estimator for $H_{ij}^{\mathrm{inhom}}(r)$ as 
\begin{align}\label{eq:Hest}
    \widehat{H}_{ij}^{\mathrm{inhom}}(r)
    =
    1
    -
    \left(
    \sum_{x \in X_i \cap W_{\ominus r}} \frac{1}{\lambda_i(x)}
    \right)^{-1}
    \sum_{x \in X_i \cap W_{\ominus r}}
    \frac{1}{\lambda_i(x)}
    \prod_{y \in X_j \cap b(x,r)}
    \left[
    1- \frac{\inf \lambda_j}{\lambda_j (y)}
    \right],
\end{align}
which was originally presented by \cite{Cronie2016}; $b(x,r)$ stands for a disc of radius $r$ centered at $x$, and $W_{\ominus r}$ is an $r$-reduced window defined as 
\begin{align*}
    W_{\ominus r} 
    =
    \left\{
    u \in W: d(u, \partial W ) \geq r
    \right\},
\end{align*}
where $\partial W$ is the border of window $W$. An estimator for $F_{j}^{\mathrm{inhom}}(r)$ is given as 
\begin{align}\label{eq:Fest}
    \widehat{F}_{j}^{\mathrm{inhom}}(r)
    =
    1
    -
    \frac{1}{
    |\mathcal{G} \cap W_{\ominus r}|
    }
    \sum_{u \in \mathcal{G} \cap W_{\ominus r}}
    \left[
    \prod_{x \in X_j \cap b(u,r)} 
    \left[
    1- \frac{\inf \lambda_j}{\lambda_j (y)}
    \right]
    \right],
\end{align}
where $\mathcal{G}$ is a fine grid defined over $W$ \citep{van11}. Having these two estimators, one can estimate the cross-type $J$-function. We add that, in practical situations, one can make use of function \textsf{Jinhom.cross} from the \textsf{R} package \textsf{spatstat} \citep{Baddeley2015}. It's worth noting that the inhomogeneous $J$-function extends beyond pairwise interactions, and it might be more precise than the $K$-function in certain scenarios. In fact, as pointed out by DS, in the example of influenza virus proteins, the cross-type $J$-function seems to deviate from the theoretical value for a marked Poisson process quicker than the cross-type $K$-function. This might have happened due to the construction of $J$-functions going beyond pairwise interactions, cf. \citet[Figure 1]{van11}. 
The performance of non-parametric estimators for $K$- and $J$-functions may also play a role.

Looking at the discussed estimators, it is evident that we must first estimate the intensity functions. Note that in real scenarios, we do not have access to the true intensity functions, and thus, one needs to estimate them in advance. We add that based on our experience with at least $K$-functions for point processes on planar spaces, simulation studies have revealed that utilizing the true intensity functions in the estimators of $K$-functions results in a reduction in performance.  Kernel estimation undeniably stands out as the primary method for non-parametric intensity estimation; this was also highlighted by MCvL. Within the literature, for a given unmarked point pattern $\x = \{ x_1,x_2,\ldots, x_n \}$ the most frequently used kernel-based intensity estimators are
\begin{equation}
\label{e:kde.2D.unif}
\widehat \lambda^{\text{U}}_{\sigma}(u)
= 
\frac{1}{c_{\sigma,W}(u)} \sum_{i=1}^n \mathcal{K}_{\sigma}(u - x_i),
\quad u \in W,
\end{equation}
and
\begin{equation}
\label{e:kde.2D.JD}
\widehat \lambda^{\text{JD}}_{\sigma}(u)
= 
\sum_{i=1}^n \frac{\mathcal{K}_{\sigma}(u - x_i)}{c_{\sigma,W}(x_i)},
\quad u \in W,
\end{equation}
where \(\mathcal{K}_{\sigma}\) is a symmetric  density function on \(\mathbb R^2\) with bandwidth \(\sigma\), and
\begin{equation}
\label{e:cW}
c_{\sigma,W}(u)
= 
\int_W \mathcal{K}_{\sigma}(u - v) \mathrm{d}v,
\quad u \in W,
\end{equation}
is the mass of the kernel centred at \(u \in W\), playing the role of an edge corrector to compensate for the lack of information outside \(W\). The estimator \eqref{e:kde.2D.unif}, which is unbiased if the true intensity is constant, is often called \textsf{uniformly-corrected}, and \eqref{e:kde.2D.JD}, which conserves mass, is called \textsf{Jones-Diggle} \citep[Chapter 6]{Baddeley2015}. In situations where the observed window is not regular, according to \cite{baddeley2022diffusion}, these two estimators may suffer from tunnelling mass as well as simultaneous under and over-smoothing. Thus, \cite{baddeley2022diffusion} proposed a kernel-based intensity estimator where intensity is estimated via a transition probability density of a Brownian motion on $W$ that respects a boundary. Their estimator is given as
\begin{equation}\label{eq:heat}
\widehat \lambda_t (u)
=
\sum\limits_{i=1}^n 
\mathcal{K}_t (u|x_i)
,
\end{equation}
where \(t= \sigma^2\), \(\sigma\) is the smoothing bandwidth, and \(\mathcal{K}_t (\cdot|x_i)\) is the heat kernel. The performance of the above intensity estimators heavily depends on the smoothing bandwidth, according to which a small bandwidth leads to low bias and high variance, whereas a large bandwidth yields high bias and low variance. Different bandwidth estimators, based on different perspectives, are proposed within the literature: Diggle's approach \cite{diggle1985kernel}, likelihood cross-validation \citep{loader2006local}, Scott's rule-of-thumb \citep{scott2015multivariate}, and Cronie and van Lieshout's criterion \citep{cronie2018non}. Since a single smoothing bandwidth may not lead to an estimate which performs equally well all over the observed window, adaptive kernel-based intensity estimators are proposed, which use a set of locally estimated bandwidth \citep{abramson1982bandwidth, davies2018fast, rakshit2019fast, van2021infill}.   Alternative to kernel-based intensity estimators are Voronoi-based estimators, which are adapted to the local variability of the underlying point process and generally outperform kernel-based estimators; see \citet{barr2010voronoi, MoradiVor2019, MATEU2020Pseudo-separable} for details. Nevertheless, according to our experience, kernel-based estimators with less variability give rise to more reliable estimates of the summary statistics; \cite{cronie2020inhomogeneous} and \cite{MoradiSharifi} used kernel-based intensity estimators with Scott's rule-of-thumb when estimating $J$-functions. A comprehensive practical review of non-parametric intensity estimators based on real-data scenarios, with all examples being reproducible, is presented by \cite{MateuMoradinonparamint}.

Turning to mark correlation functions, the numerator of the $\tf_f$-correlation function
\begin{eqnarray}\label{eq:tfcorr}
\kappa_{\tf_f}(r)
    =
    \frac{
    \ee \left[
    \tf_f \left(
    m(x),m(y)
    \right) \big| x,y \in X
    \right]
    }{
    c_{\tf_f}
    },
    \quad  
    d(x,y)=r,
\end{eqnarray}
is estimated via
\begin{align}\label{eq:mcorrest}
    \frac{
    \sum_{x,y \in X}^{\neq}
    \tf_f 
    \left(
    m(x),m(y)
    \right)
    \mathcal{K}(
    d(x,y) - r
    )
    w(x,y)
    }{
    \sum_{x,y \in X}^{\neq}
     \mathcal{K}(
    d(x,y) - r
    )
    w(x,y)
    },
\end{align}
where $ \mathcal{K}$ is a kernel function on the real line and $w(x,y)$ is an edge correction factor \citep[Chapter 15]{Baddeley2015}; the normalization factor $c_{\tf_f}$ is the sample average of $\tf_f (m(x),m(y))$ taken over all $x,y \in X$. It's often noted in the literature that these estimators can be computed without edge corrections, particularly when both the numerator and denominator are estimated using the same principle \citep[Chapter 5]{Illian2008}. In fact, in our article, such as in Figure 6, no edge-correction was applied in computing Stoyan's mark correlation function.

As for marked point processes on linear networks, we point to \cite{Baddeley2014} for non-parametric estimators of second-order cross/dot-type summary characteristics. Non-parametric estimators for mark-weighted $K$-functions could be defined by closely following \citet{Ang2012, rakshit2017second}. Regarding cross/dot-type higher-order summary characteristics such as $J$-functions, their non-parametric estimators could be achieved by following \citet{Cronie2016, cronie2020inhomogeneous}. Finally, non-parametric estimators of mark correlation functions would be of a similar nature as in \eqref{eq:mcorrest} but with a distance metric adapted to linear networks. We emphasise that the choice of distance metric might very well depend on the application under study.

\section{Marked spatial point process models}

In our article, we restricted our focus to the current state-of-the-art for mark summary characteristics. CJK and MRS commented on lacking a thorough discussion on point process models when points are marked. In particular, MRS provided a profound overview of the matter, which we greatly appreciate.   
Indeed, the literature covers various doubly-stochastic models such as Cox point processes, where the intensity itself is driven by a random field, and Markov point process models, which formalize the intensity as a product over clique potentials, i.e. energy functions. It would be welcomed to provide an in-depth treatment of the current state-of-the-art of models for marked spatial point processes, including all proposed estimation methods. Given the thorough and compact discussion of the subject given by MRS, we here only point to some further models and recent extensions.

Notable early contributions for marked point process models include the balanced and linked Cox models of \cite{https://doi.org/10.1111/j.2517-6161.1983.tb01224.x} in which a bivariate point process $\boldsymbol{X}=\lbrace X_1, X_2 \rbrace$ is driven by a non-negative valued bivariate random field $\boldsymbol{Z}=\lbrace Z_1, Z_2\rbrace$ such that given the realizations $\lambda_1(u)=Z_1(u)$ and $\lambda_2(u)=Z_1(u)$ for all $u\in \R^2$ the components of $\boldsymbol{X}$ are  independent inhomogeneous
Poisson processes with intensity functions $\lambda_i(\cdot),\ i=1,2$. For any $\nu>0$, the points of the two components $X_1, X_2$ may then show clustering/repulsion tendencies with $Z_1=\nu Z_2$ in the case of the linked Cox process and with $Z_1+Z_2=\nu$ for the balanced Cox process model, respectively. Less restrictive extensions of the Cox model include the marked versions of the log-Gaussian Cox process \citep{lgcp} and the shot-noise Cox process \citep{BrixShortNoise}. 
Although log-Gaussian Cox processes appear most commonly in the literature, we point to the work of \cite{https://doi.org/10.1111/biom.12339} who proposed a multivariate product-shot-noise Cox process to model a multispecies point pattern. Allowing also for temporal variation of spatial marked processes, various spatio-temporal models can be considered, including growth-interaction, spatial birth-and-death and Hawkes processes; these were also highlighted by CJK. Note also a combination of log-Gaussian Cox processes with Hawkes processes, i.e. the so-called Cox-Hawkes process, proposed by  \cite{miscouridou2022coxhawkes}, which could also be extended to marked cases.        
Moreover, we also point to the Candy \citep{Stoica2004,Candy} and Bisous \citep{TEMPEL201617} models, which, instead of points model line segment and, thus, might be useful tools to derive models for (marked) spatial movement data.

Lastly, we appreciated the idea of MRS treating the mark sum as a local mark characteristic. Indeed, for a fixed observed point $x$, the normalized mark-sum measure adjusts the number of points in a disc $b(x,r)$ by the sum of the marks in $b(x,r)$ and can, thus, be interpreted as the contribution of marks in a distance $r$ centred at $x$ \citep{Pettinen:2007}.

\section{Applications of mark summary characteristics}

We are grateful to DS for proposing the computation of different mark correlation functions for some given data and discussing their abilities to describe mark correlations. It is important to highlight that any such mark summary characteristic addresses a particular property of the mark distribution.
To illustrate the application of different mark correlation functions, we revisit the three considered models I to III of Section 4.2.2 in our article. In a similar manner, for each model, we simulate 199 patterns for which we calculate Stoyan's mark correlation function $\kappa_{mm}^{\LL}(r_{\LL})$, Beisbart and Kerscher’s mark correlation function $\kappa_{mm}^{\mathrm{Bei},\LL}(r_{\LL})$, mark variogram $\gamma_{mm}^{\LL}(r_{\LL})$, and Shimantani's $I$ function $I_{mm}^{\mathrm{Shi},\LL}(r_{\LL})$. We used the notation $\kappa$ for, e.g. Stoyan's mark correlation function, to be better distinguished from the cross/dot-type $K$-functions.
Based on the 199 estimated mark correlation functions per each test function, we obtain $95\%$ pointwise critical envelopes to compare the general behaviours of the four chosen test functions; one could instead use global envelopes \citep{mari1, MrkvickaEtal2020} as pointed out by MRS. Concerning DS's comment on the differences between the mark variogram and (semi-)variogram, although both are constructed through half-squared distances, they are in general different except under the geostatistical marking model where marks are generated from an underlying random field \citep{Schlather2004, GUAN2007148}. In this model, also referred to as a random field model due to its construction, the marks and the locations are generally assumed to be independent. We note that the geostatistical marking model can be tested against deviations from the independence assumption between marks and points, i.e. non-geostatistical marking model, by e.g. Schlather's $E(r)$ and $V(r)$ functions and the tests proposed in \citep{10.1111/j.1541-0420.2005.00395.x}

\begin{figure}[!h]
    \centering
    \includegraphics[scale=0.12]{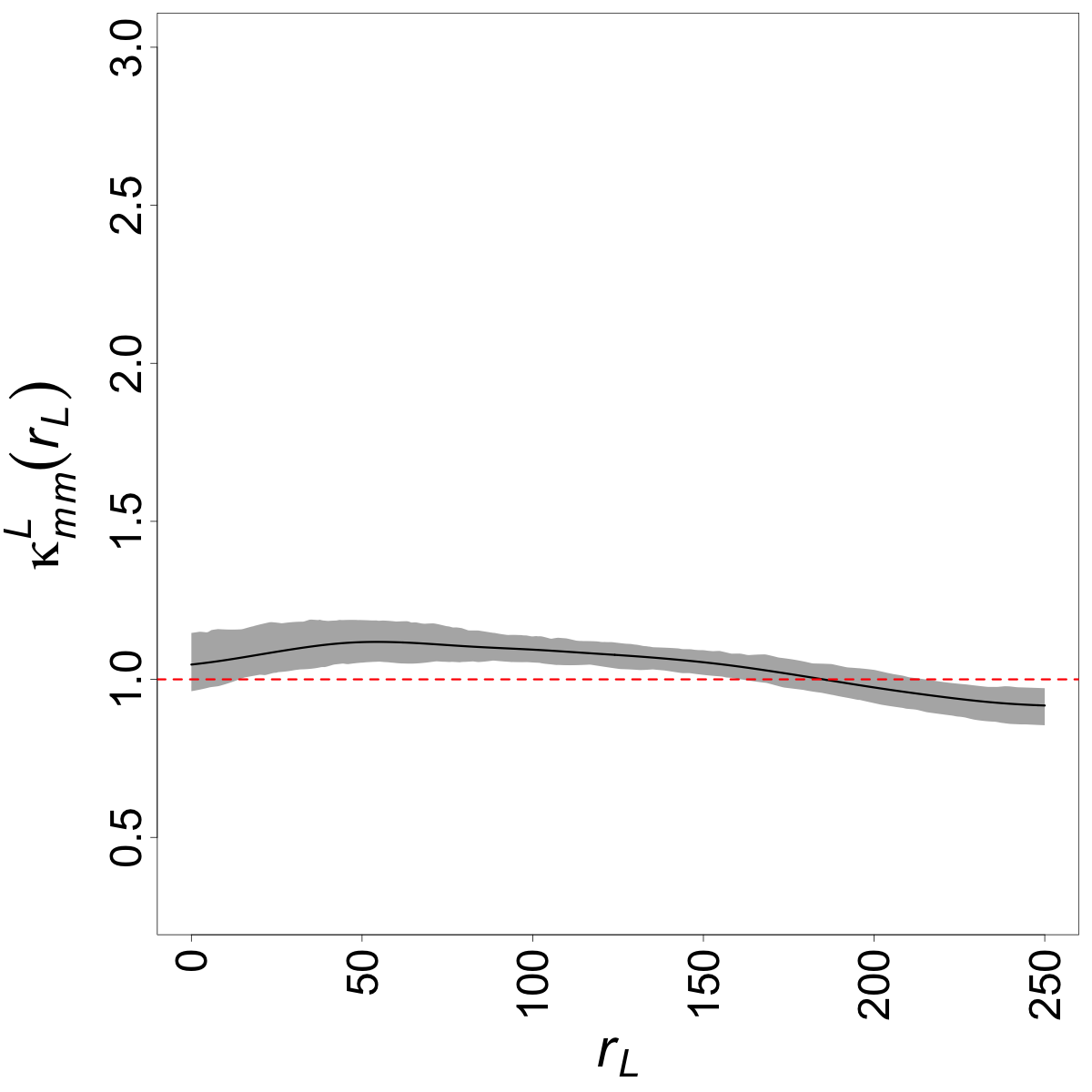}
    \includegraphics[scale=0.12]{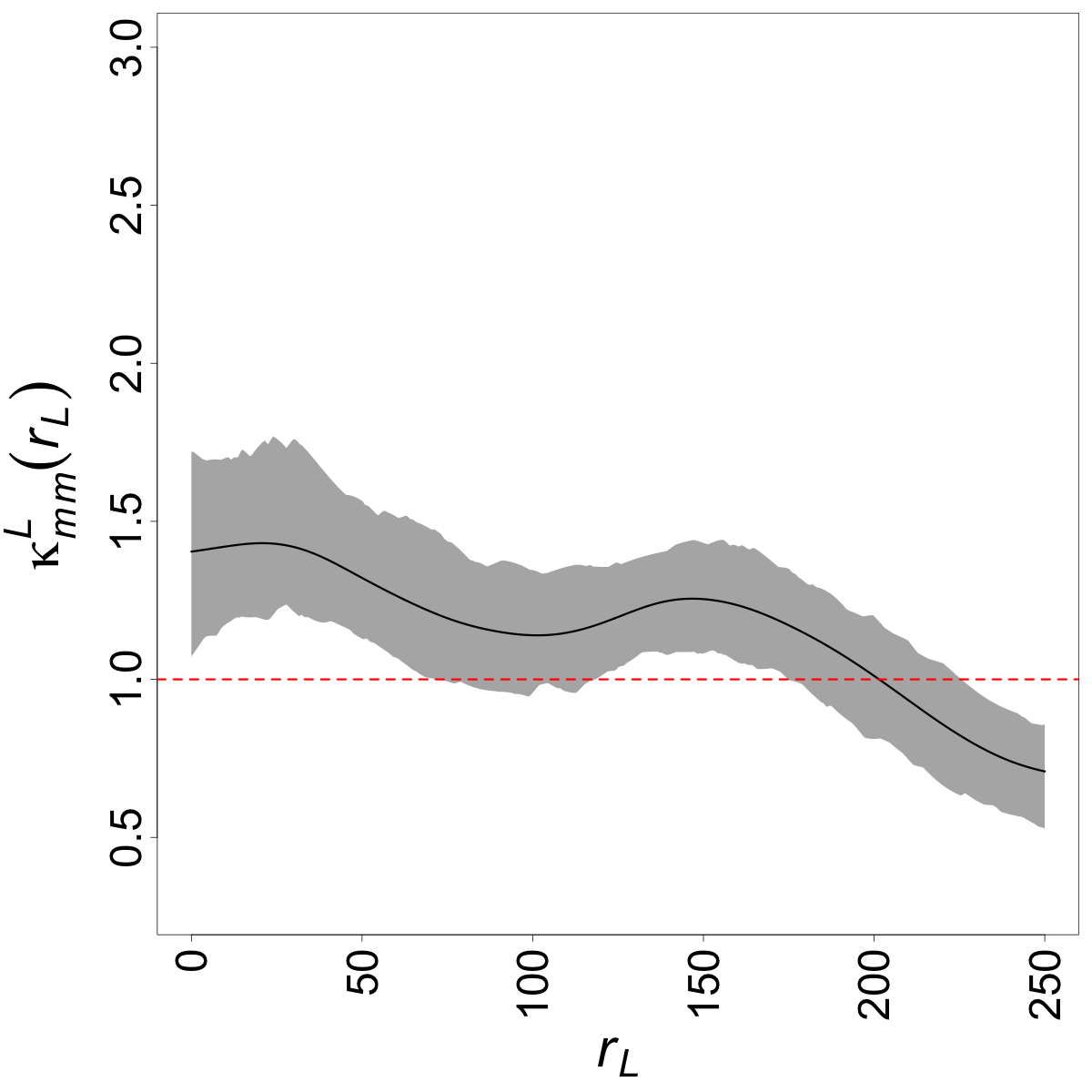}
    \includegraphics[scale=0.12]{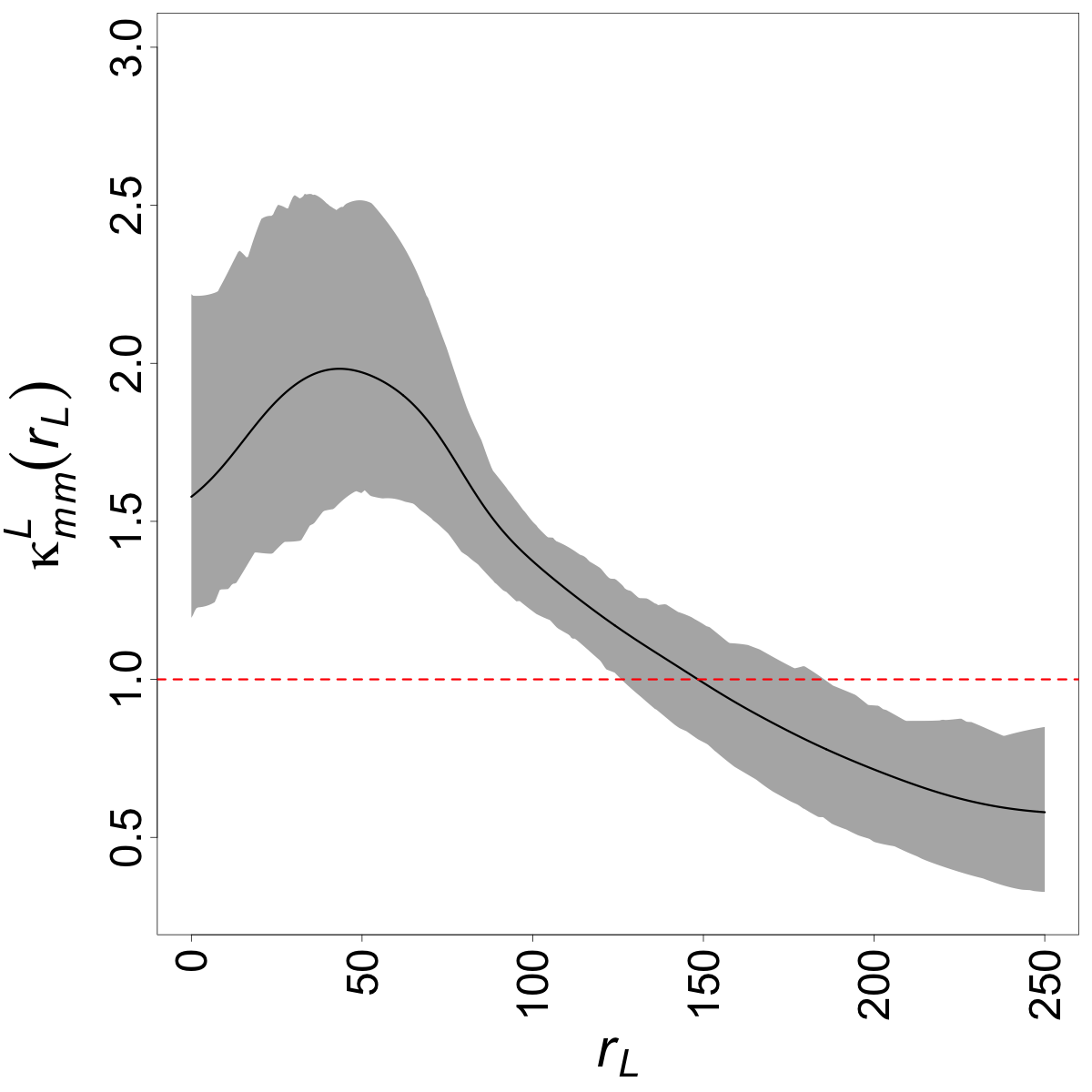}
    \includegraphics[scale=0.12]{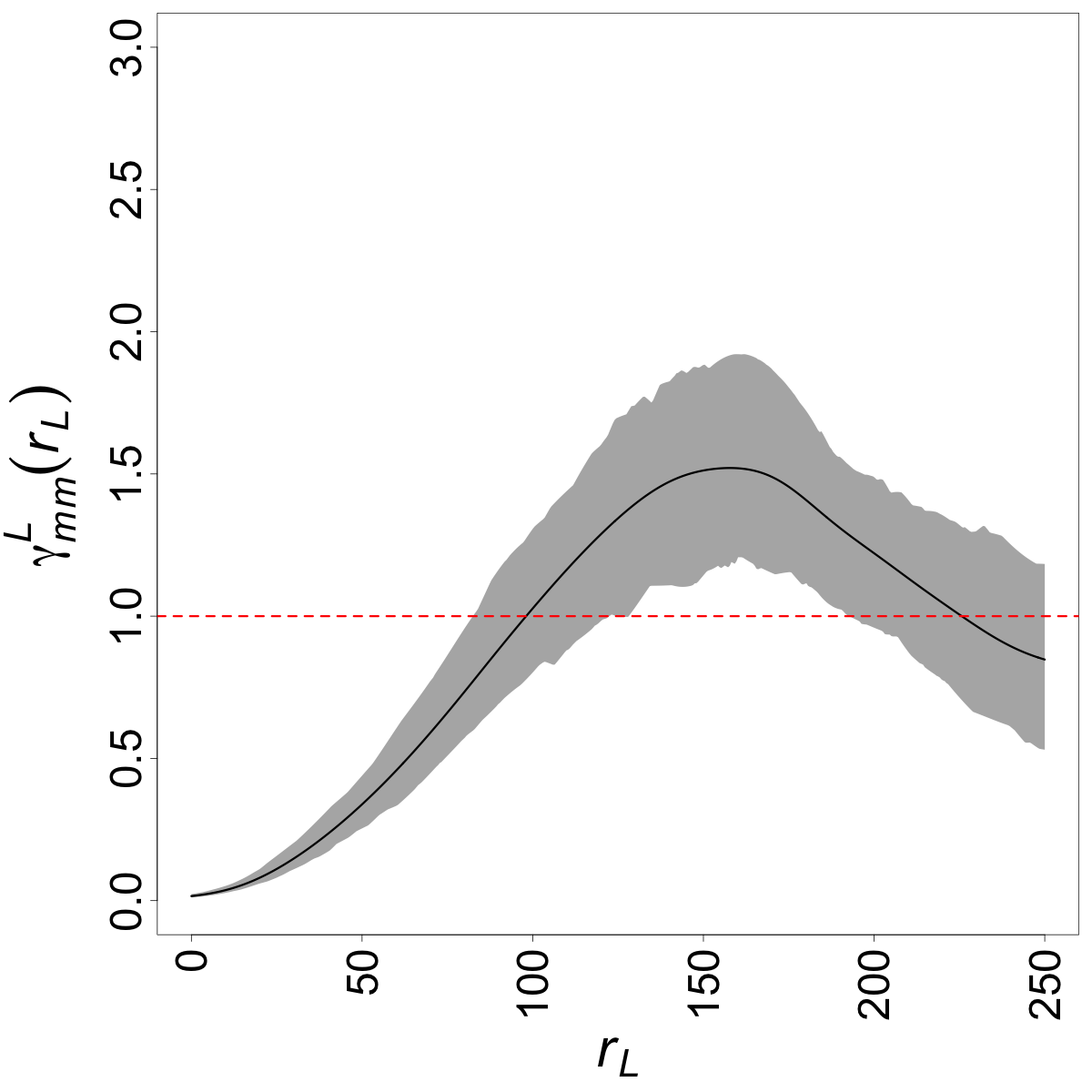}
    \includegraphics[scale=0.12]{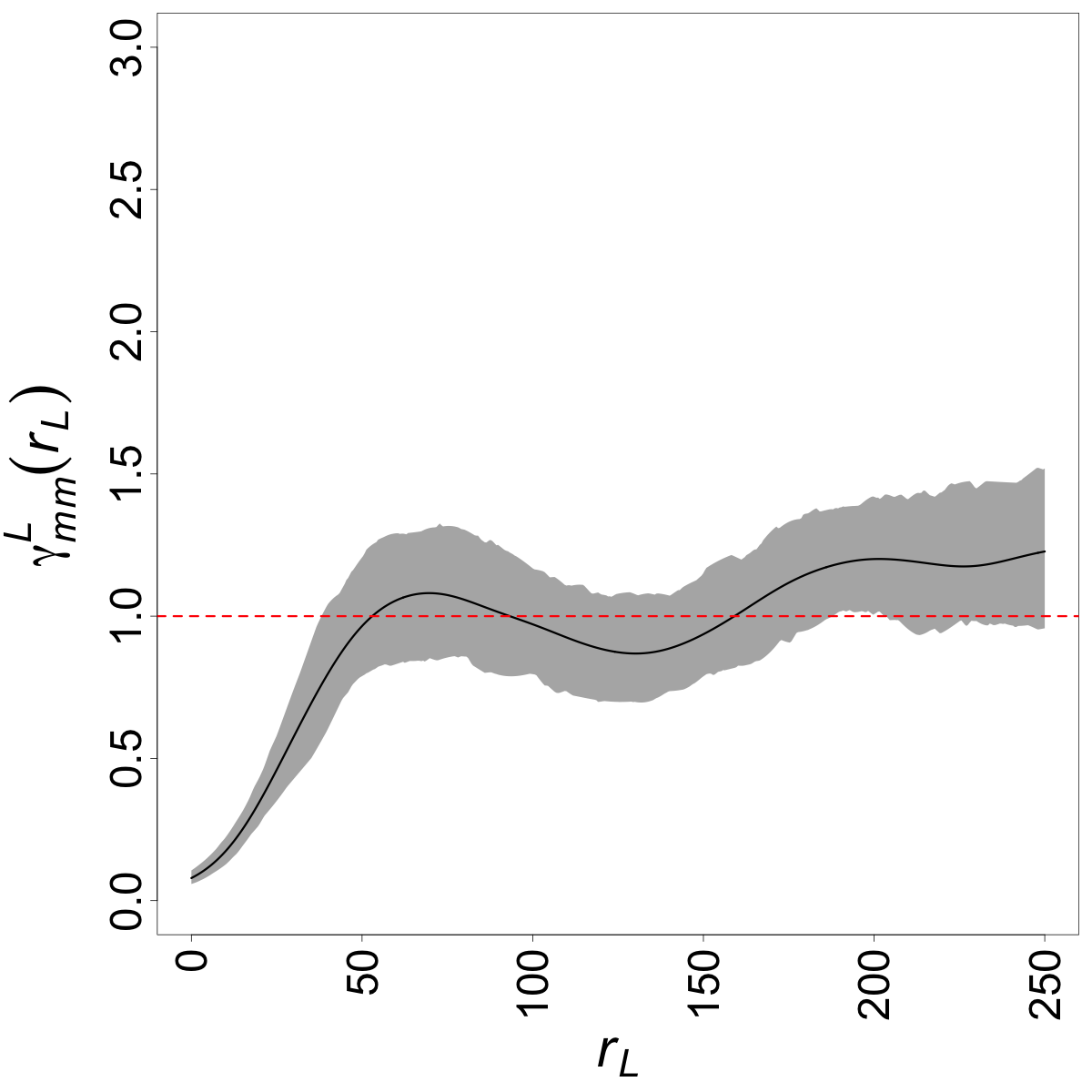}
    \includegraphics[scale=0.12]{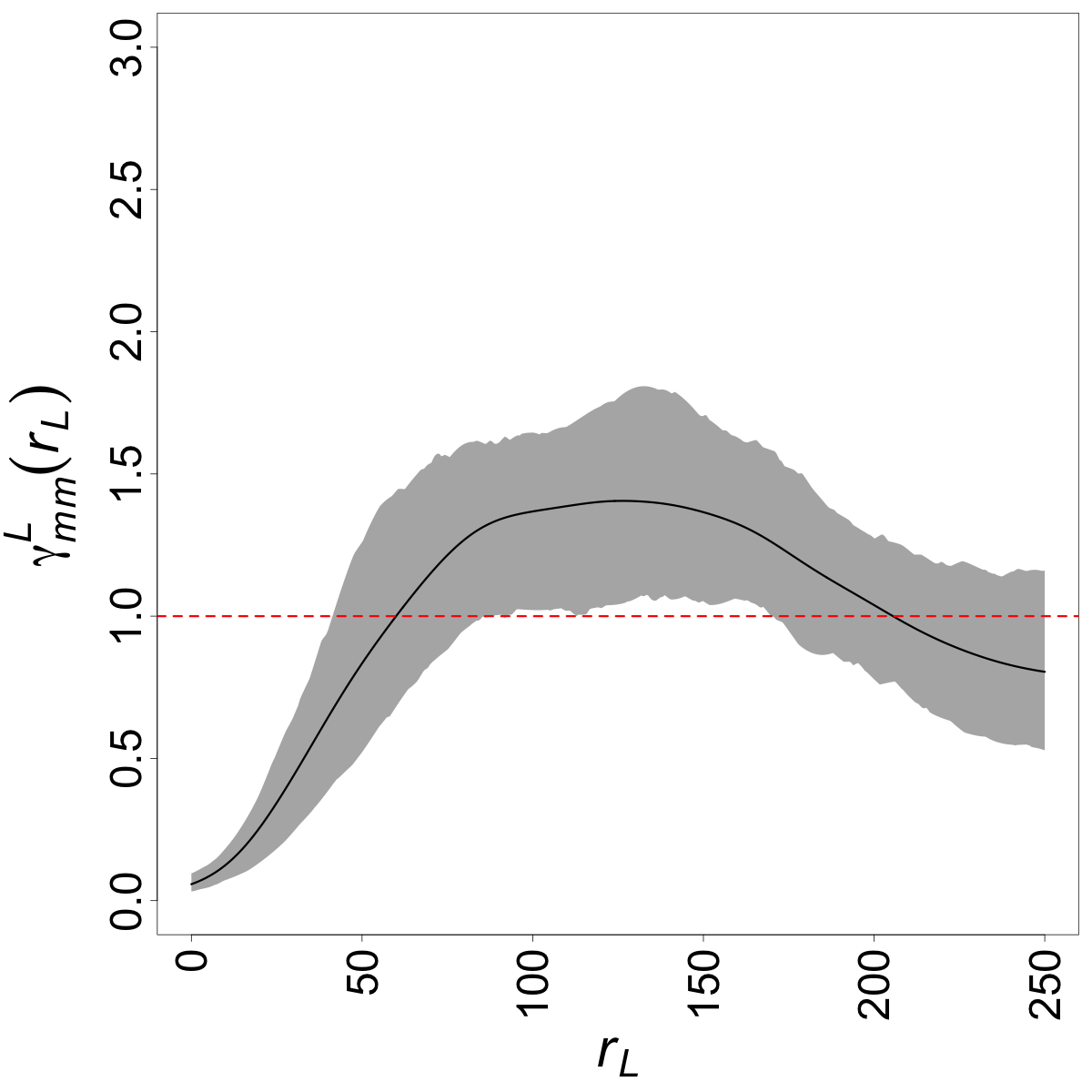}
    \includegraphics[scale=0.12]{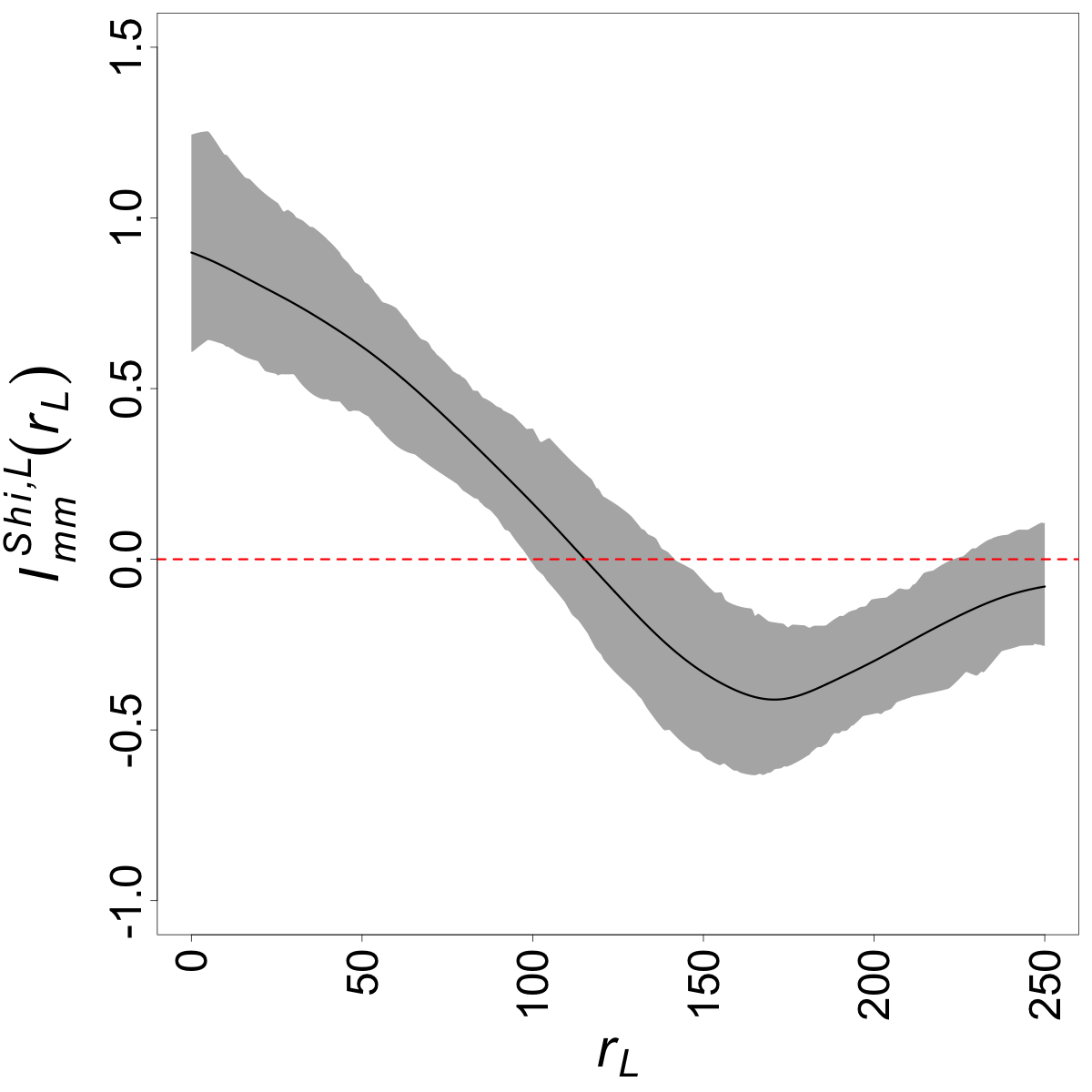}
    \includegraphics[scale=0.12]{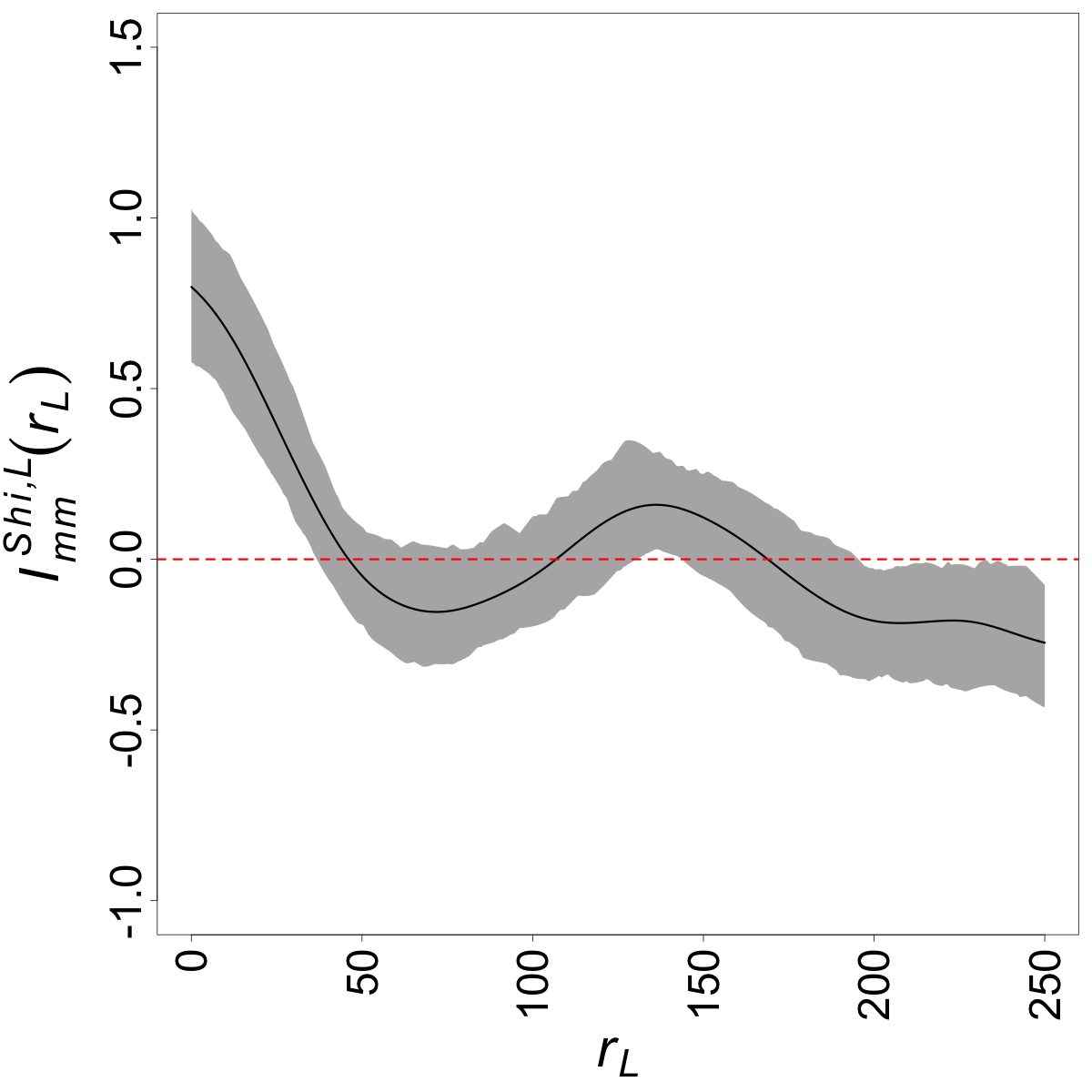}
    \includegraphics[scale=0.12]{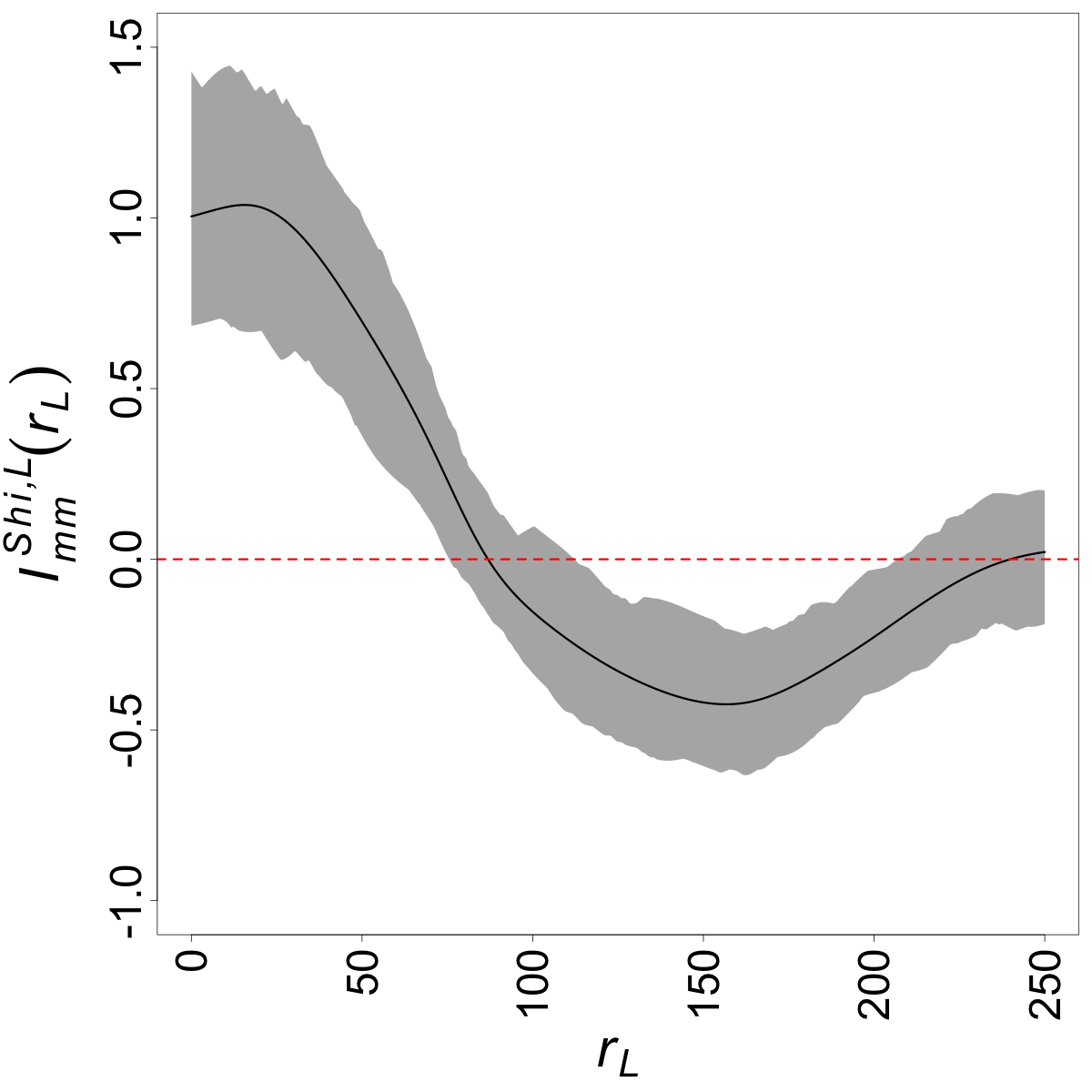}
    \includegraphics[scale=0.12]{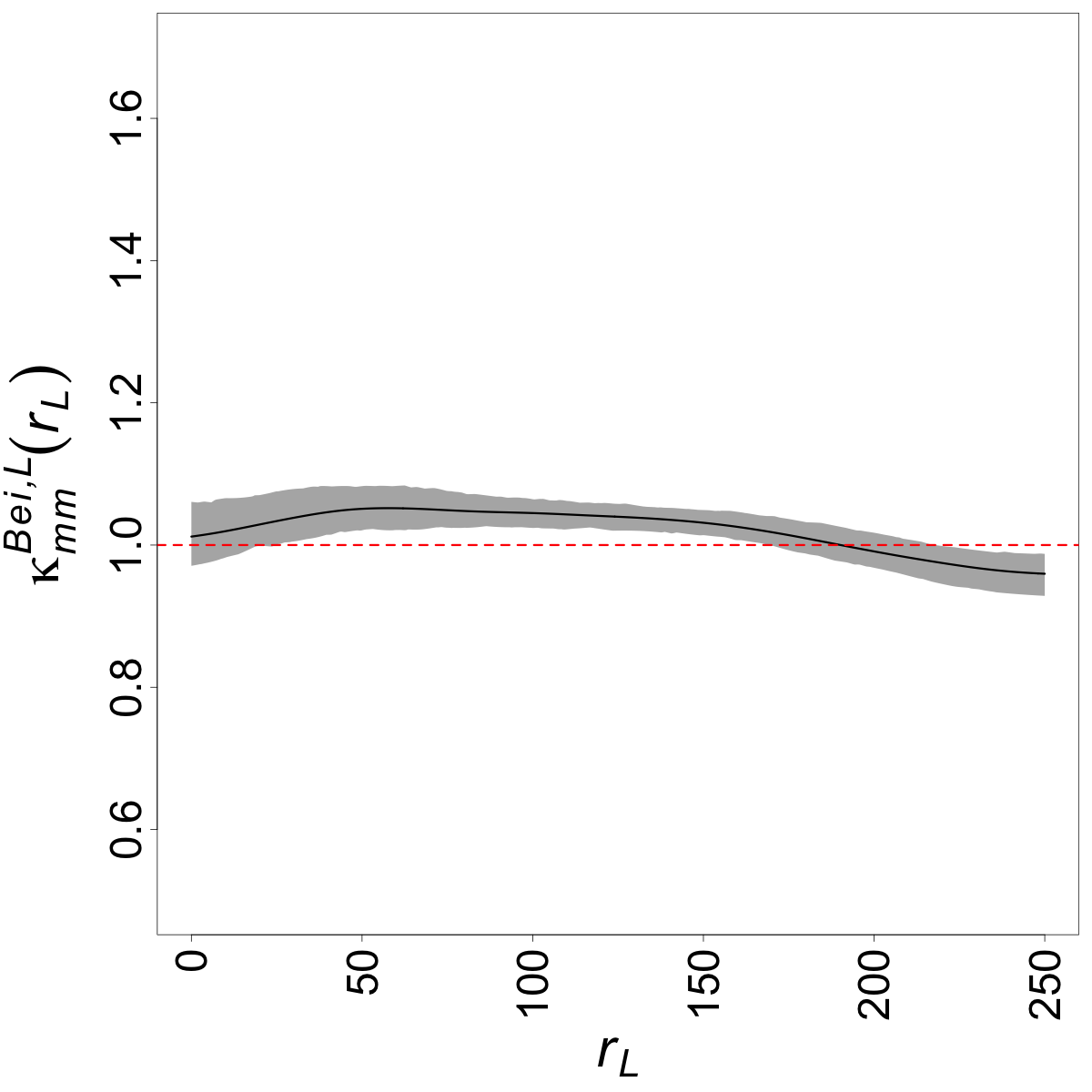}
    \includegraphics[scale=0.12]{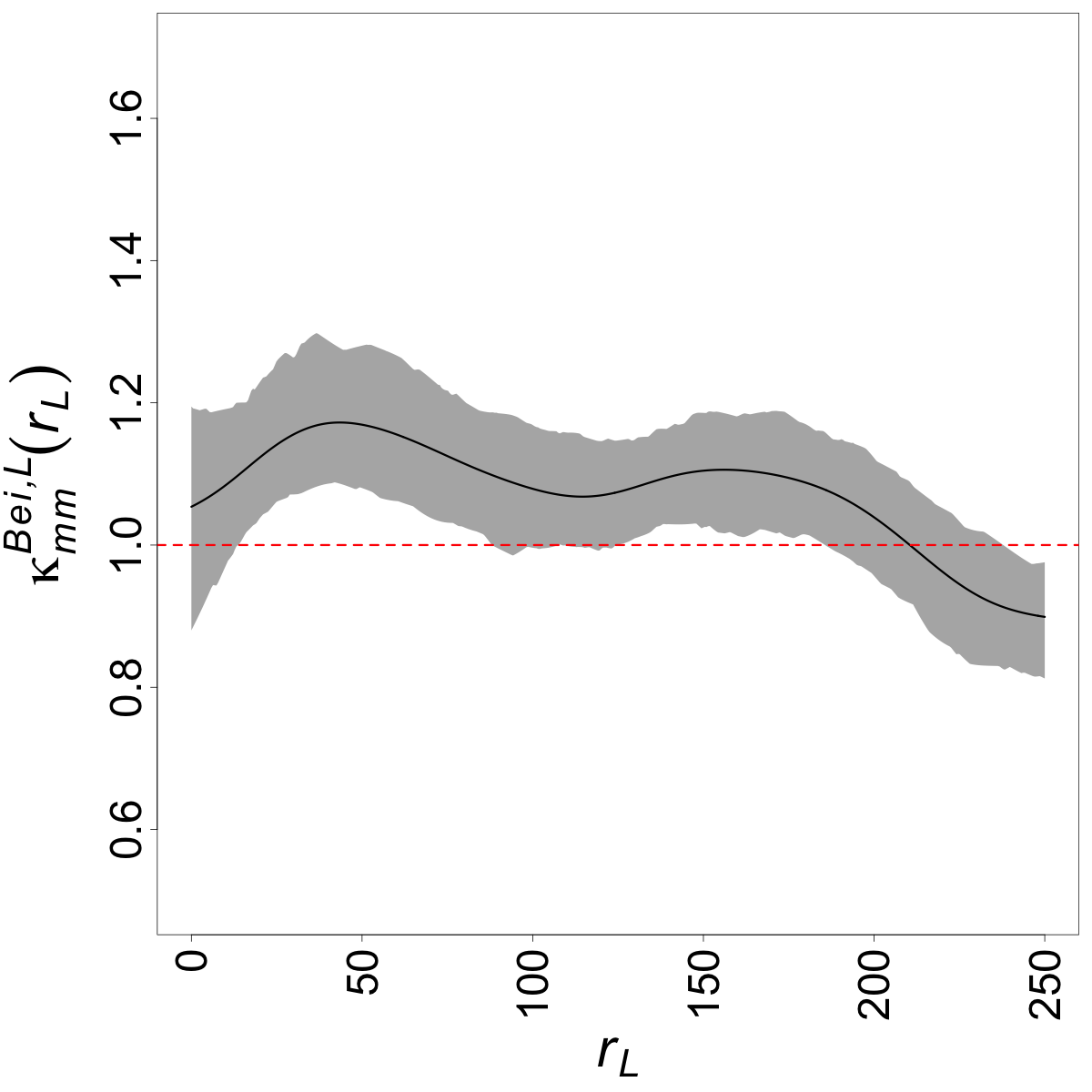}
    \includegraphics[scale=0.12]{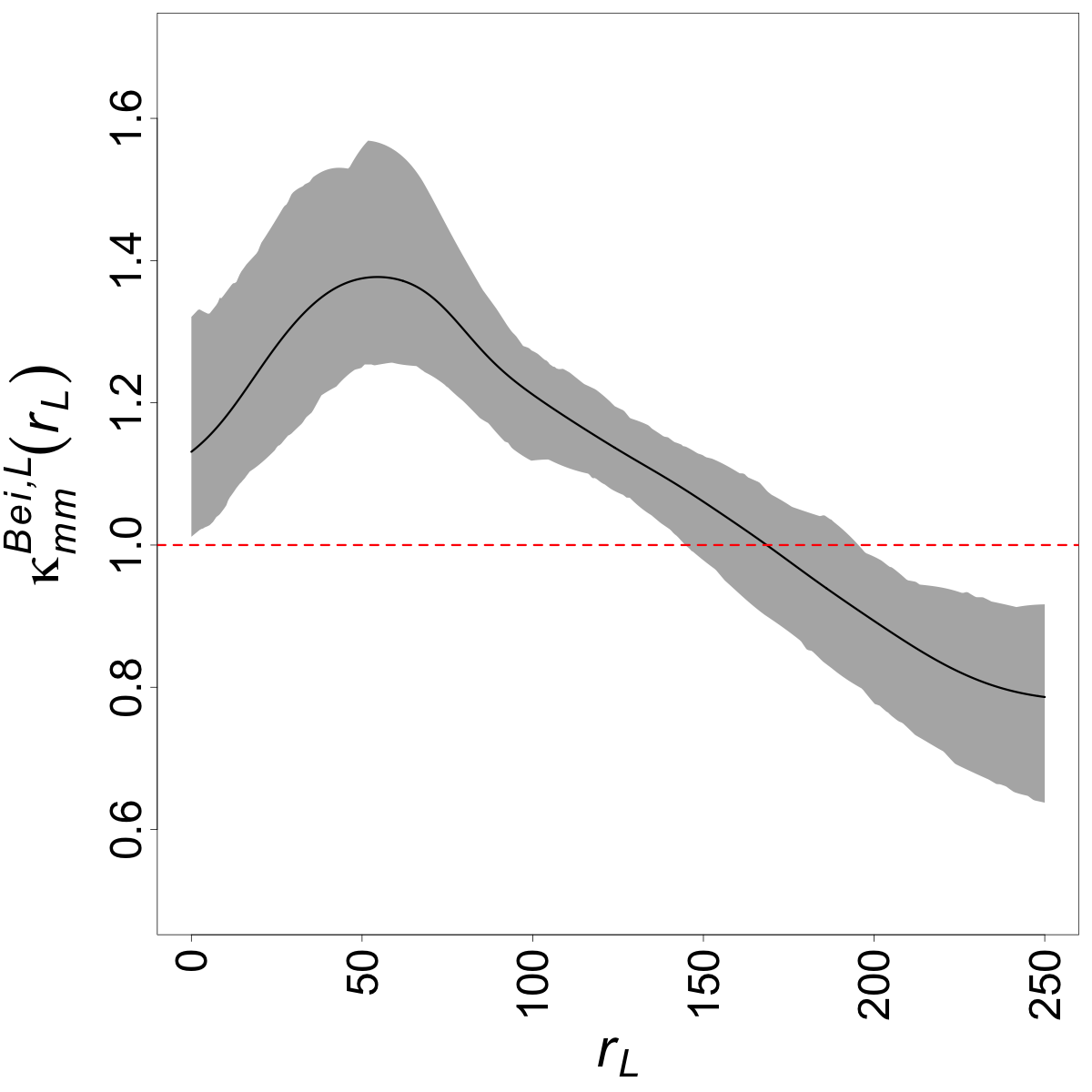}
    \caption{
    $95\%$ critical envelops for the Stoyan's mark correlation function $\kappa^{\LL}_{mm}(\rL)$, Mark variogram  $\gamma^{\LL}_{mm}(\rL)$, Shimanti's $I^{\mathrm{Shi},\LL}_{mm}(\rL)$, and Beisbart and Kerscher's mark correlation function $\kappa^{\mathrm{Bei},\LL}_{mm}(\rL)$, and their averages, based on 199 simulated patterns from Models I, II, and III, from left to right, respectively.
    }
    \label{fig:additionalsim}
\end{figure}

Turning to some further numerical evaluations, for model I (left column of Figure \ref{fig:additionalsim}), in which the point configuration inherits a trend from bottom-left to top-right of the network, for nearby points (small values of $r_{\LL}$), we can see low variation (based on mark variogram $\gamma^{\LL}_{mm}(\rL)$) with some positive association (based on Stoyan's mark correlation function $\kappa^{\LL}_{mm}(\rL)$ and Beisbart and Kerscher's mark correlation function $\kappa^{\mathrm{Bei},\LL}_{mm}(\rL)$) and high positive spatial autocorrelation (based on Shimanti's $I^{\mathrm{Shi},\LL}_{mm}(\rL)$). With increasing distance $r_{\LL}$, the marks become more heterogeneous, yielding a clear increase in spatial variation as it can be seen from the mark variogram $\gamma^{\LL}_{mm}(\rL)$, less association indicated by Stoyan's mark correlation function $\kappa_{mm}^{\LL}(r_{\LL})$ and Beisbart and Kerscher’s mark correlation function $\kappa_{mm}^{\mathrm{Bei},\LL}(r_{\LL})$ and, correspondingly, negative spatial autocorrelation revealed by Shimantani's $I$ function $I_{mm}^{\mathrm{Shi},\LL}(r_{\LL})$. 
For a more thorough exploration of how these mark correlation functions differ in their effectiveness at describing mark correlations,
we add that if pairs of points have similar mark values, the variation between the marks is small, leading to small values of the mark variogram $\gamma^{\LL}_{mm}(\rL)$ close to zero, while the association is high, yielding values above one for Stoyan's mark correlation function $\kappa^{\LL}_{mm}(\rL)$ and Beisbart and Kerscher's mark correlation function $\kappa^{\mathrm{Bei},\LL}_{mm}(\rL)$. In cases where the marks for any pair of points, deviating from the mark mean, demonstrate substantial similarity (or dissimilarity) in value, positive (negative) autocorrelation is expected to be revealed by Shimanti's $I^{\mathrm{Shi},\LL}_{mm}(\rL)$. While the focus here is limited to point patterns on linear networks, we emphasise that these interpretations hold for point patterns on $\R^2$ as well.


Turning to the central column of Figure \ref{fig:additionalsim}, which shows the results for model II, we again notice a distinct opposite behavior between the mark variogram $\gamma^{\LL}_{mm}(\rL)$ and the mark correlation functions $\kappa^{\LL}_{mm}(\rL)$ and $\kappa^{\mathrm{Bei},\LL}_{mm}(\rL)$, especially at short distances.
Recalling the construction of the marks in model II based on the shortest-path distance of each point to the dendrite’s border, nearby points are expected to have similar marks. As the distance $r_{\LL}$ increases, the marks become more dissimilar, which yields a strong variation in value and, at the same time,  low association. However, the marks at very long distances, say at two distinct points at different borders, are again more similar due to the simulation design. Here, we would point out the fact that such long shortest-path distances are not captured in any of the four plots due to the restriction to  $r_{ \mathds{L},\mathrm{max}} = 250$. 

Similar results are also obtained for model III (right column of Figure \ref{fig:additionalsim}), in which the marks correspond to the number of nearest-neighbors which are at a network distance less than $\rL < 80$ units from a target point. As nearby neighbors are expected to have similar mark values, the mark variogram $\gamma^{\LL}_{mm}(\rL)$ at short distances possesses values close to zero. A high similarity is also indicated by the mark correlation functions $\kappa^{\LL}_{mm}(\rL)$ and  $\kappa^{\mathrm{Bei},\LL}_{mm}(\rL)$ having values larger than one and Shimantani's $I$ function having positive values. If the interpoint distance becomes large, heterogeneity in the mark values increases, yielding an increase in the mark variation and a decrease in mark association and autocorrelation. Note that, due to the construction of marks in model III, the mark association is expected to be maximized around $\rL=80$, which is properly detected by $\kappa^{\LL}_{mm}(\rL)$ and $\kappa^{\mathrm{Bei},\LL}_{mm}(\rL)$. Reinspecting Figure 3 in our article, the highest mark values appear in the very central parts of the dendrite, while the lowest mark values occur near the borders of the dendrite; note the lack of information near borders. Having this in mind, the similarity and, thus, the spatial autocorrelation between marks is high, and consequently, the mark variation is low when the interpoint distance $\rL$ is small; this is particularly emphasized in the central region of the dendrite.

\medskip

Furthermore, we thank DS for pointing to the interpretation of the results concerning the public street trees in Vancouver, Canada. In the network setting, it is important to note that the interpretation of the results becomes more challenging and requires a solid investigation of both the network structure and the phenomena under study. It might be the case that the shortest-path distance is not the most suitable metric for this application. However, note that, in this application, although trees are located along the Vancouver street network, and the characteristics are computed using the shortest-path distance, the trees might be influenced by similar soil conditions, which, in turn, might affect their growth, shapes or canopies. Note that there might be trees that are spatially close to each other while they possess a large shortest-path distance, e.g., trees on two nearby disconnected streets. 
Thus, careful consideration of these potential external influences is crucial when interpreting the results. Specifically, for biological, ecological, and environmental data happening on network structures, it's vital not to underestimate these effects and to evaluate them while considering both the network's specificity and expertise knowledge on the phenomena under study. Unfortunately, we did not have access to any further information, which would have allowed for a more critical evaluation of the findings. Taking a more detailed look at the age of the public street trees, which was only reported for a very small proportion of trees, we did not identify any significant structure in the age distribution for this subset of trees except for a slight indication of heterogeneity in the trees' ages.

\section*{Acknowledgement}

The authors also gratefully acknowledge financial support through the German Research Association and the Stochastic Group of the German Mathematical Society. Matthias Eckardt was funded by the Walter Benjamin grant 467634837 from the German Research Foundation. Mehdi Moradi received travel support from the Stochastic Group of the German Mathematical Society.

\section*{Declarations}

The authors have no conflicts of interest to declare.

\bibliography{Rejoinder}
\bibliographystyle{chicago}

\end{document}